\documentclass[]{aastex}
\usepackage{graphics,emulateapj5}
\newcommand{\bfx}{{\bf x}}
\newcommand{\bfv}{{\bf v}}
\newcommand{\bfq}{{\bf q}}
\newcommand{\pin}{{\sc PINOCCHIO}}
\def \gtsima{\mbox{$\; \buildrel > \over \sim \;$}}
\shorttitle{Formation and evolution of dark matter haloes}
\shortauthors{Monaco et al}

\begin{document}
\title{Predicting the number, spatial distribution and merging history 
of dark matter haloes}
\author {Pierluigi Monaco\altaffilmark{1}, Tom Theuns\altaffilmark{2}, Giuliano Taffoni\altaffilmark{3}, 
Fabio Governato\altaffilmark{4}, Tom Quinn\altaffilmark{5} and Joachim Stadel\altaffilmark{5}}
\altaffiltext{1} {Dipartimento di Astronomia,   Universit\`a di Trieste, 
via Tiepolo 11, 34131 Trieste, Italy}
\altaffiltext{2} {Institute of Astronomy, Madingley Road, Cambridge CB3 0HA, UK}
\altaffiltext{3} {SISSA, via Beirut 4, 34014 Trieste, Italy}
\altaffiltext{4} {Osservatorio Astronomico di Brera, Merate, Italy}
\altaffiltext{5} {Astronomy Department, University of Washington, Seattle, USA}

\begin{abstract}
We present a new algorithm 
(PINOCCHIO, PINpointing Orbit-Crossing Collapsed HIerarchical objects)
to predict accurately the formation and
evolution of individual dark matter haloes in a given realization of an
initial linear density field. Compared with the halo population formed
in a large ($360^3$ particles) collisionless simulation of a CDM
universe, our method is able to predict to better than 10 per cent
statistical quantities such as the mass function, two-point correlation
function and progenitor mass function of the haloes. Masses of
individual haloes are estimated accurately as well, with errors
typically of order 30 per cent in the mass range well resolved by the
numerical simulation. These results show that the hierarchical
formation of dark matter haloes can be accurately predicted using local
approximations to the dynamics when the correlations in the initial
density field are properly taken into account. The approach allows one
to automatically generate a large ensemble of accurate merging
histories of haloes with complete knowledge of their spatial
distribution. The construction of the full merger tree for a $256^3$
realisation requires a few hours of CPU-time on a personal computer,
orders of magnitude faster than the corresponding $N$-body simulation
would take, and without needing any extensive post-processing.  The
technique can be efficiently used, for instance, for generating the input 
for galaxy formation modeling.
\end{abstract}


\section{Introduction}
In currently favoured dark matter dominated cosmological models,
initially small density fluctuations are amplified by gravity and
eventually condense out of the Hubble expansion to form gravitationally
bound systems at a density contrast of $\gtsima 200$ -- dark matter
haloes (e.g. Peebles 1993).  The properties of the halo population are
of fundamental importance for understanding galaxy formation and
evolution. Indeed, galaxies are thought to form when baryons fall into
such dark matter haloes and are shocked to sufficiently high
temperatures and densities that the gas can cool radiatively to form
stars (Rees \& Ostriker 1977, White \& Rees 1978).

The formation of haloes can be studied using numerical simulations
which usually evolve a set of equal mass particles that represent the
dark matter in a periodic simulation box (e.g. Efstathiou et
al. 1985). A popular way of identifying \lq haloes\rq\ in such
calculations is the friends-of-friends (FOF) algorithm, which links
particles within a fraction $b$ of the mean inter-particle spacing into
one halo, at a density contrast of $\gtsima 1/b^3$. Other halo
identification algorithms generally give similar results. Jenkins et
al. (2001) combined the results from simulations with a variety of box
sizes to obtain the mass function $n(M)$ of FOF haloes over a large
dynamic range.

Analytical descriptions of the halo formation process were pioneered
by Press \& Schechter (1974, hereafter PS) and were recently reviewed
by Monaco (1998). Although the PS mass function and its extensions
(the so-called excursion set approach, Bond et al. 1991) fit the
numerical FOF mass function reasonably well (e.g. Efstathiou et
al. 1988), there are real discrepancies both at large and small masses
where PS respectively under and over predicts halo numbers
(e.g. Governato et al. 1999; Jenkins et al. 2001; Bode et al. 2001).
Similar discrepancies are found when reproducing the mass function of
the progenitors of halos of given mass (Sheth \& Lemson 1999;
Somerville et al. 2000).  In addition, Bond et al. (1991) and White
(1996) demonstrated that the PS approach achieves a very poor
agreement on an object-by-object basis when compared to simulations
(but see Sheth, Mo \& Tormen 1999 for a different view).  Analytic
approaches based on the peaks of the initial density fields did not
achieve a better agreement with simulations (Katz, Quinn \& Gelb
1993).  Intermediate between simulations and analytical techniques are
perturbative approaches that describe the growth of haloes in a {\em
given} numerical realisation of a linear density field, such as the
truncated Zel'dovich (1970) approximation (Borgani, Coles \&
Moscardini 1994), the peak-patch algorithm (Bond \& Myers 1996a,b) and
the merging cell model (Rodriguez \& Thomas 1996; Lanzoni, Mamon \&
Guiderdoni 2000).

In this paper we present a new algorithm to compute the formation and
evolution of dark matter haloes in a given linear density field.  
A 1D version of this algorithm was given by Monaco \& Murante (2000).  
In common with the other perturbative approaches, we combine a local
description of the dynamics in order to identify collapsed haloes with
Lagrangian perturbation theory to displace the haloes to their final
positions. We demonstrate that the algorithm leads to an accurate
description of the detailed clustering and merger history of haloes
while requiring several orders of magnitude less computer time and
post-simulation analysis than the corresponding full blown numerical
simulation. In addition, the successful reproduction of the merger
history demonstrates that we have identified the key processes that
govern halo formation, and that these can be described with a
perturbative approach.

We follow a two step procedure that mimics the hierarchical built-up of
haloes through accretion and merging. The first step identifies
orbit-crossing (hereafter OC) as the instant at which a mass element
undergoes collapse. We compute OC numerically by applying local
ellipsoidal collapse approximation to the full Lagrangian perturbative
expansion (Bond \& Myers 1996a, Monaco 1995, 1997). This part is the
more computationally expensive, requiring several hours of computer
time on a personal computer
for a 256$^3$ realization. 
The second step groups the collapsed
particles into disjoint haloes, using an algorithm similar to that used
to identify haloes in $N$-body simulations.  Basically, a particle
accretes onto a halo if it is sufficiently close to it at its collapse
instant. We use Lagrangian perturbation theory (LPT, Catelan 1995,
Bouchet 1996; Buchert 1996) to compute the positions of haloes and
particles. Seed haloes are local maxima of the collapse redshift. This
second step automatically determines the full merger history of haloes
and requires negligible computer time. Compared to simulations, the
first step determines when a simulation particle enters a high-density
region whereas the second identifies the haloes.

Since our method describes, in the linear density field, the hierarchical
built-up of objects that have undergone OC, we refer to it as \pin:
{\sc PIN}pointing {\sc O}rbit-{\sc C}rossed {\sc C}ollapsed {\sc
HI}erachical {\sc O}bjects. In the next section we describe the
algorithm in more detail. In 
section 3
we compare its
predictions with those from simulations and discuss possible
applications of the method.
Section 4 gives the conclusions.  Technical details and resolution
issues are addressed in forthcoming papers (Monaco, Theuns \& Taffoni
2001; Taffoni, Monaco \& Theuns 2001).

\section{The algorithm}

\begin{figure*}
\label{fig:mfunc}
\setlength{\unitlength}{1cm}
\centering
\begin{picture}(7,9)
\put(-3, -2.5){\includegraphics{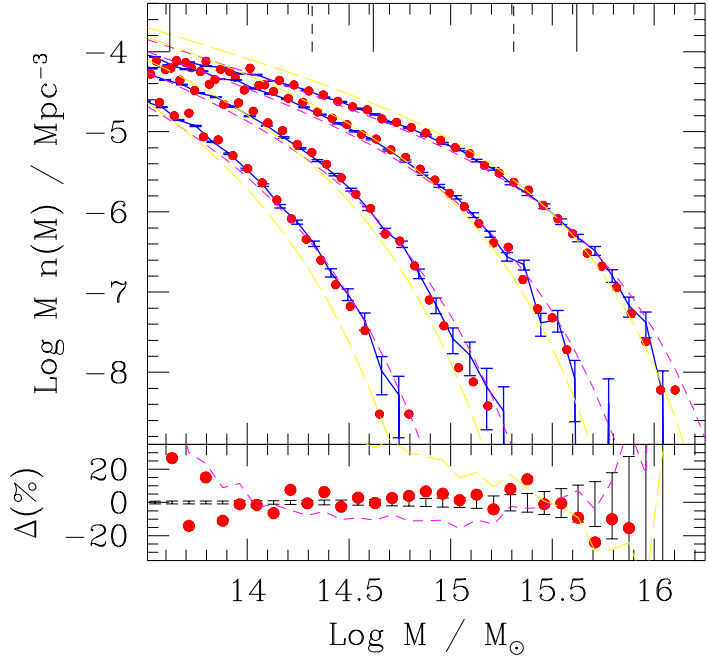}}
\end{picture}
\caption{Comparison of mass function $M n(M)$ 
in a standard CDM model ($\Omega_m=1$). Top panel: simulated mass
function for FOF selected haloes (Full lines with Poissonian error
bars), \pin\ mass function (filled circles), the fit by Sheth and
Tormen (short-dashed lines) and PS function (long-dashed lines), at
redshifts $z=0$, 0.43, 1.13 and 1.86 (higher redshift curves are
off-set by 0.1 dex both vertically and horizontally for improved
clarity).  Vertical lines show limits corresponding to simulation
haloes with 10, 50, 100, 500 and 1000 particles (256$^3$ re-sampling).
Bottom panel: Difference between simulated mass function and \pin\
(filled dots), Sheth and Tormen fit (short-dashed line) and PS
(long-dashed line) at $z=0$.}
\end{figure*}

\begin{figure*}
\label{fig:dplot}
\setlength{\unitlength}{1cm}
\centering
\begin{picture}(7,7)
\put(-2,-2){\includegraphics{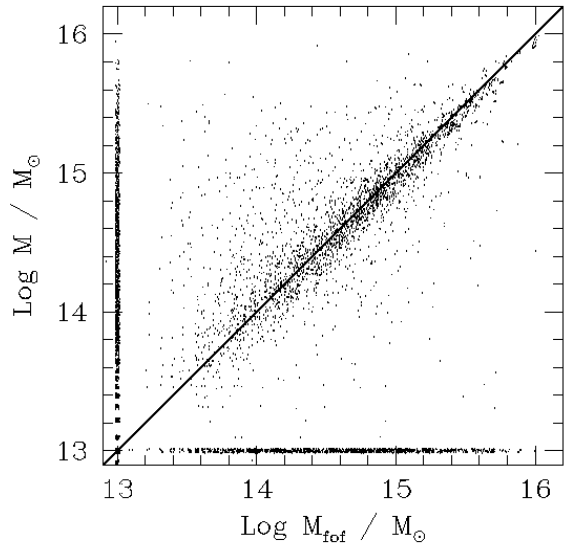}}
\end{picture}
\caption{Predicted halo mass against FOF halo mass for a subset of the
particles of Governato et al. (1999) $\Omega_m=1$ simulation at
redshift $z=0$ The \pin\ masses are highly correlated with the FOF
masses. Points which have not collapsed have been arbitrarily assigned
a mass of $10^{13}M_\odot$.}
\end{figure*}

\subsection{Orbit crossing}
Consider a random realization of a density field, $\rho(\bfq)$, where
$\bfq$ denote Lagrangian (initial) coordinates, and let $\phi({\bfq})$
be the corresponding peculiar potential. Both fields can be smoothed
by convolving them with a Gaussian with FWHM $R$; we denote them as
$\rho(\bfq,R)$ and $\phi({\bfq},R)$ respectively. The first derivative
of the potential, $\partial_{q_i}\phi$, describes the motion of the
particle in the Zel'dovich (1970) approximation, and the shear tensor,
$\partial_{q_i}\partial_{q_j}\phi$, can be used to give a description
of the deformation of the mass element based on ellipsoidal collapse
(Bond \& Myers 1996a, Monaco 1995, 1997).  In our context ellipsoidal
collapse is a convenient truncation of LPT (Monaco 1997).

For a given smoothing radius $R$, the density of a mass element will
become infinite as soon as at least one of the ellipsoid's axes reaches
zero size, at which point the relation ${\bf x}({\bf q})$ becomes
multi-valued and the Jacobian of the transformation ${\bf q}\rightarrow
{\bf x}$, $J={\rm det} |\partial \bfx/\partial \bfq|=0$. This is the
definition of OC.  We argue that after this instant $t_c$, non-linear
processes will become important and hence further predictions of what
happens to the mass element can not be safely made using LPT. However,
as the density of the mass element is already very high, we regard it
as a candidate for the building up of a collapsed halo at time
$t_c(R)$.  A different definition of collapse was used, e.g., by 
Audit, Teyssier \& Alimi (1997),  
Lee \& Shandarin (1998) and Sheth \& Tormen (2000).

In practice we generate the density field $\rho$ on a cubic grid. In
our description, mass elements (or \lq particles\rq) then correspond to
the grid vertices ${\bf q}$. The potential $\phi$ and its derivatives
are computed from $\rho$ using fast Fourier transforms. We typically
use $\sim $20 logarithmically spaced smoothing radii. Applying local
ellipsoidal collapse to each particle, we obtain the collapse redshift
on each smoothing scale, and we record for each particle the highest
collapse redshift $z_c$, the corresponding smoothing scale $R_c$ and
the Zel'dovich (1970) estimates for the peculiar velocity
$\bfv_c(R_c)\propto\nabla\phi({\bf q},R_c)$ on that smoothing scale.
Note that at this stage we make no prediction of the {\em mass} of the
collapsed halo that the particle accreted onto.

In fact, the collapsed mass element will not necessarly have accreted
onto any halo, but may instead have become part of a filament or sheet
(collectively referred to as \lq filaments\rq\ hereafter), since these
have undergone OC as well. These structures trace the moderate over
densities that connect the much higher density collapsed haloes in
simulations. The next subsection describes how the OC region is divided
into collapsed haloes and OC filaments.

\subsection{Fragmentation}
The grouping of OC particles into haloes mimics the hierarchical
formation of objects, and also the way in which halo finders identify
collapsed objects in simulations. We begin by sorting particles
according to decreasing collapse redshift $z_c$, and, starting from the
highest $z_c$ we decide the fate of the collapsed particle, working our
way down forward in time to the last particle to collapse.

Briefly, at the instant the particle is deemed to collapse, we decide
which halo, if any, it accreted onto.  The candidate haloes are those
that already contain one {\em Lagrangian neighbour} of the
particle\footnote{On the initial grid ${\bf q}$ of Lagrangian
positions, the six particles nearest to a given particle are its \lq
Lagrangian neighbours\rq.}.  The particle will accrete onto the halo if
it is \lq sufficiently close\rq\ to it at the collapse time, mimicking
the construction of FOF haloes. We use the Zel'dovich velocities ${\bf
v_c}$ as defined earlier to compute the distance, at the collapse time,
between the particle and the candidate halo.  If a particle has more
than one candidate halo, we also check whether these haloes should
merge, using a similar merger criterion.  Notice that in this way
haloes are by construction connected regions in Lagrangian space.

More in detail, we apply the following rules for accretion and
merging. (Lengths are in units of the grid spacing; $R_{\rm M}=M^{1/3}$
is the \lq radius\rq\ of a halo of $M$ particles.)

\begin{enumerate}
\item [(1)] {\em Seed haloes} Local maxima of the collapse
redshift $z_c$ are seeds for a new halo.

\item [(2)] {\em Accretion} A collapsing particle (not a local
maximum) accretes onto a candidate halo (i.e. containing one of its
Lagrangian neighbours) if the distance $d$, at the collapse time,
between particle and halo centre-of-mass is $d \le f_{\rm a}R_{\rm
M}$.  $f_{\rm a}$ is a parameter of order unity, analogous to the
linking length parameter used to identify FOF haloes.  If the particle
is able to accrete onto two (or more) haloes, we assign it to the
one for which $d/R_{\rm M}$ is the smallest.

\item [(3)] {\em Merging} If the particle has more than one
candidate halo, then these haloes are merged if their mutual distance
$d$, again at the particle's collapse time, is $d \le f_{\rm m}R_{\rm
M}$, where $R_{\rm M}$ refers to the larger halo and $f_{\rm m}$ is
again a parameter of order unity. Since we only consider six Lagrangian
neighbours, up to six haloes may merge at a given time, although binary
and ternary
mergers are of course much more frequent.

\item [(4)] {\em Filaments} With these rules for accretion and merging,
some collapsing particles do not accrete onto a halo at their collapse
time. Since these particles tend to occur in the mildly overdense
regions that connect the haloes (visible as a filamentary network
between haloes in simulations), we assign them to a \lq filaments\rq\
group. In $N$-body simulations, some particles accrete onto a halo
directly from this filamentary network, {\em without} passing through a
collapsed halo first. In order to account for this, we check the
Lagrangian 
neighbours of a particle that accretes onto a halo according to the
accretion condition (2). If any of these neighbours already belong to
the filaments group, then they also accrete onto that halo. (So up to
five additional particles may accrete onto the halo, if their common
Lagrangian neighbour satisfies condition (2)).

\end{enumerate}

When the groups are very small, $R_{\rm M}$ is comparable to the grid
spacing, and the Zel'dovich displacements are often not sufficiently
accurate for the accretion or merging condition to be fulfilled. This
resolution effect results in producing too few small haloes at high
redshift. To remedy this we improve the accretion condition to $d <
f_{\rm a} R_{\rm M} + f_{\rm r}$, and similarly for merging. Our
algorithm thus contains three parameters ($f_{\rm a}$, $f_{\rm m}$ and
$f_{\rm r}$) which need to be calibrated using the FOF mass function as
determined from a simulation, and which have obvious physical
interpretations in terms of accretion, merging and resolution effects.
Optimal values are $f_{\rm a}=0.18$, $f_{\rm m}=0.35$ and $f_{\rm
r}=0.7$. 
These values were obtained by comparing the PINOCCHIO mass function
with those of several simulations, including the standard SCDM one
discussed below, a $\Lambda$CDM simulation ($\Omega_m=0.3$,
$\Omega_\Lambda=0.7$, $\sigma_8=1$, $h=0.7$) run with the same
simulation code and box size (500 $h^{-1}$ Mpc) and another
$\Lambda$CDM simulation ($\Omega_\Lambda=0.7$, $\Omega_m=0.3$,
$\sigma_8=0.9$, $h=0.65$) with different resolutions ($128^3$ and
$256^3$ particles) in a smaller box of 100 $h^{-1}$ Mpc, evolved with
the P3M {\sc HYDRA} code (Couchman 1991).
The agreement between \pin\ and these other simulations is as good as
the comparison with the SCDM simulation described in the next section;
the best fit parameters are found to agree within $\sim$0.01.
However,
at smaller and more non-linear scales more subtle resolution effects
appear, which can be corrected for. These details will be discussed in
the forthcoming paper Monaco et al. (2001).

\section{Results}

\begin{figure*}
\label{fig:obj}

%\plotone{paperI.obj.ps}

\setlength{\unitlength}{1cm}
\centering
\begin{picture}(7,12)
\put(-3., -4){\includegraphics{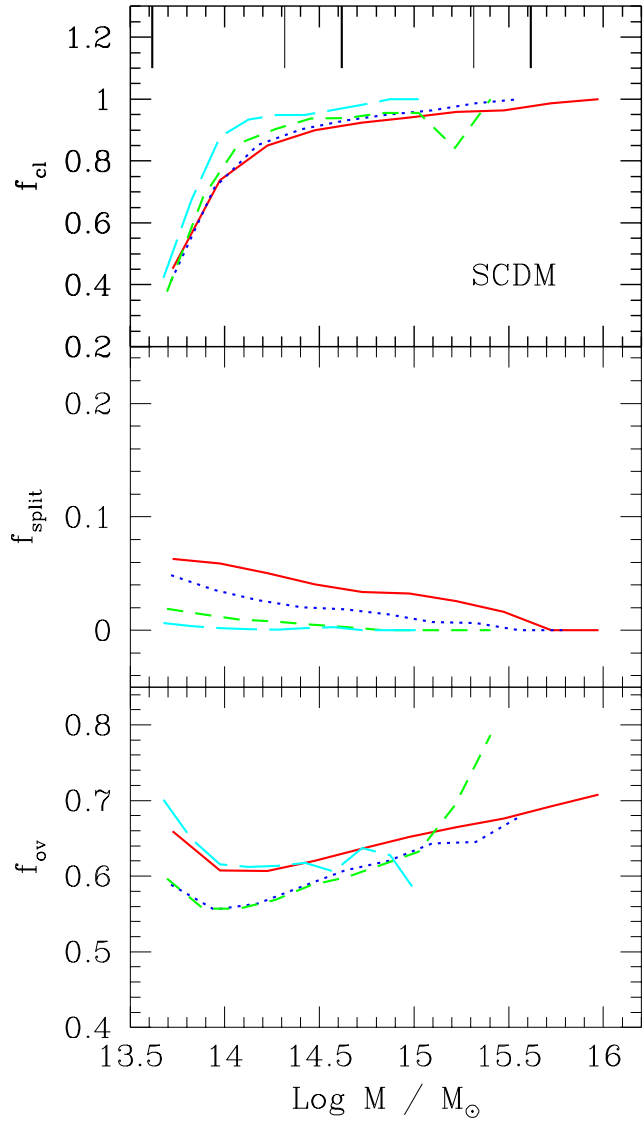}}
\end{picture}
\caption{Statistics of halo overlap between \pin\ and FOF objects for
haloes at $z=0$ (continuous lines), $z=0.43$ (dotted lines), $z=1.13$
(short dashed lines) and $z=1.86$ (long dashed lines).  Upper panel:
fraction $f_{\rm cl}$ of \lq cleanly assigned\rq\ halo pairs between
the two catalogs, as a function of mass. Middle panel: fraction
$f_{\rm split}$ of FOF halos that are split in two \pin\ halos. Lower
panel: average overlap in Lagrangian space, $f_{\rm ov}$, for cleanly
paired-up halos. (See text for definitions of $f_{\rm cl}$ and $f_{\rm
split}$.)  As in Figure 1, vertical lines show limits corresponding to
simulation haloes with 10, 50, 100, 500 and 1000 particles (256$^3$
re-sampling).}
\end{figure*}

\begin{figure*}
\label{fig:dplot2}

%\plotone{paperI.mass.ps}

\setlength{\unitlength}{1cm}
\centering
\begin{picture}(10,10)
\put(0, -1.5){\includegraphics{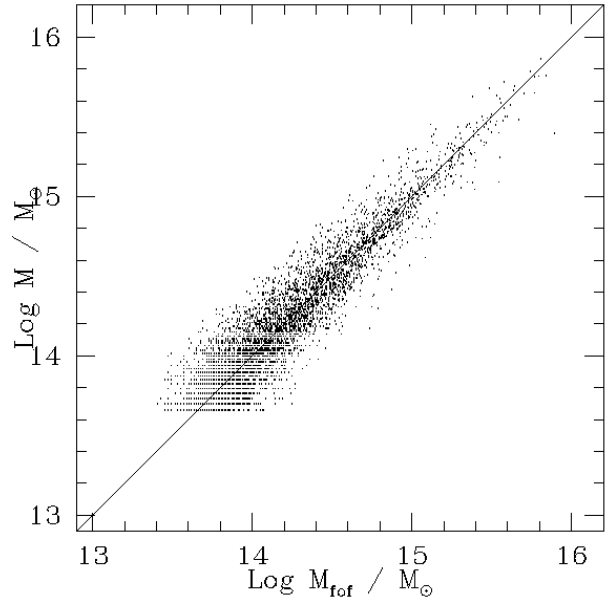}}
\end{picture}
\caption{\pin\ versus FOF halo masses for objects that were cleanly
assigned between the two respective catalogs. Note that each dot
corresponds to a halo pair, which contrasts with
Figure~\ref{fig:dplot}, where each dot refers to a {\em random} point
in the initial conditions.}
\end{figure*}

\begin{figure*}
\label{fig:prog}

%\plotone{paperI.prog.ps}

\setlength{\unitlength}{1cm}
\centering
\begin{picture}(7,9)
\put(-4.5, -5){\includegraphics{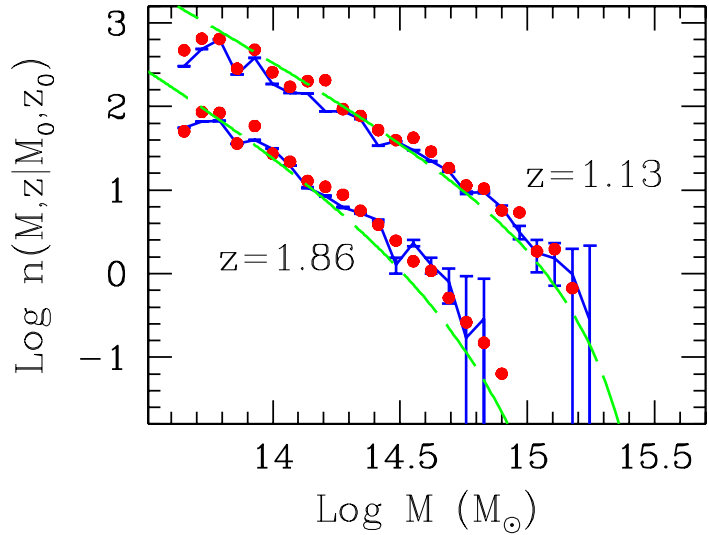}}
\end{picture}
\caption{Conditional mass function of haloes with mass $M_0=5\times
10^{15}M_\odot$ at redshift $z=0$, at the earlier redshifts 1.13 and
1.86 as indicated. Full lines with Poissonian error bars are for the
simulation, filled circles correspond to the \pin\ prediction,
long-dashed lines are the conditional mass function from the PS theory
(Bower 1991). The higher redshift results have been off-set vertically
by 1~dex for clarity. The \pin\ mass function follows the simulations
significantly better than the PS one.}
\end{figure*}

\begin{figure*}
\label{fig:corr}

%\plotone{paperI.corr.ps}

\setlength{\unitlength}{1cm}
\centering
\begin{picture}(7,11)
\put(-1.5, -4.8){\includegraphics{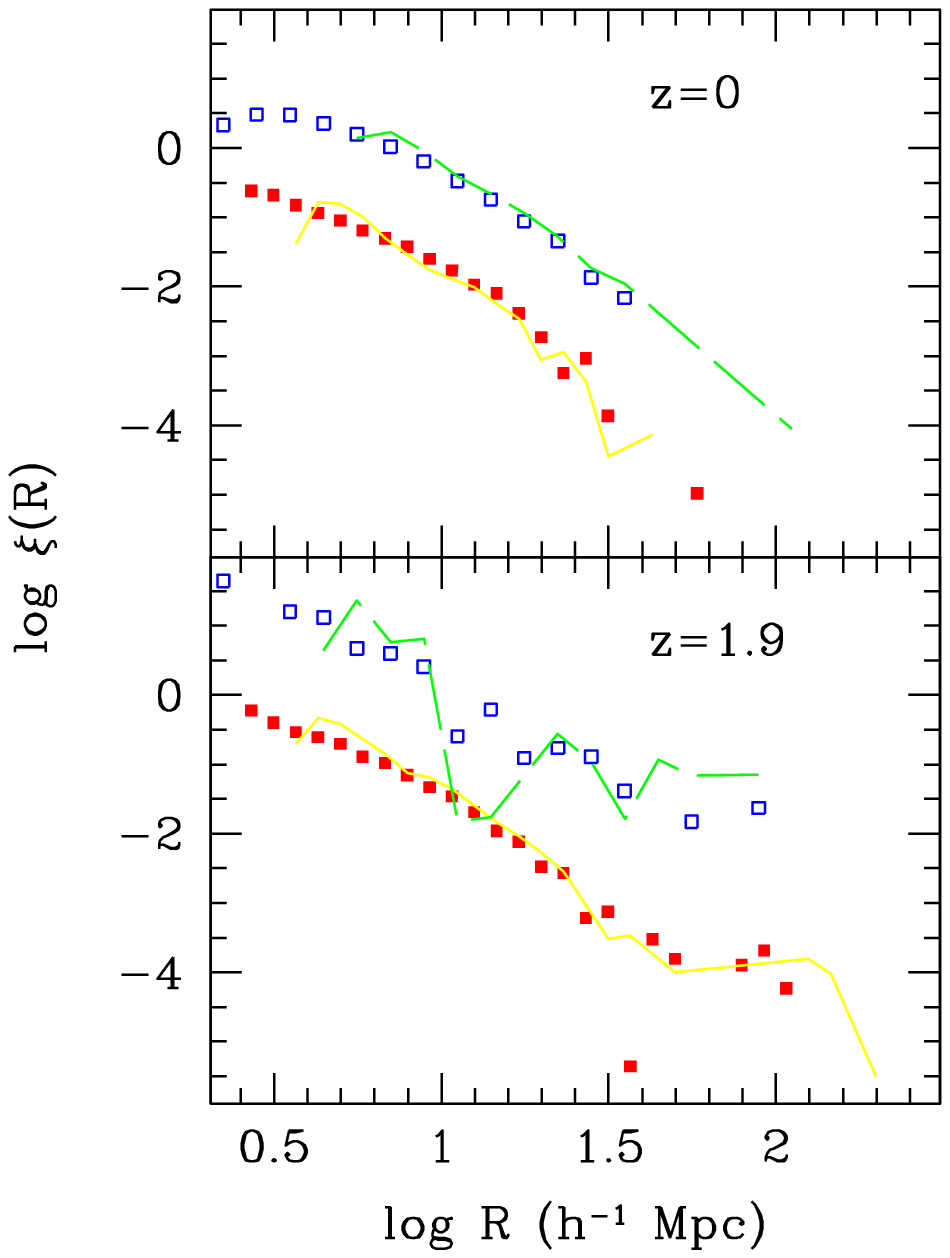}}
\end{picture}
\caption{Correlation functions for haloes within a given mass range as
a function of co-moving separation $R$, for the two redshifts indicated in
the panels. Symbols refer to FOF selected haloes, lines to \pin\
haloes. Mass ranges are $10^{14}\le \log M/M_\odot\le 10^{14.5}$
(filled squares and full lines), and $\log M/M_\odot\ge 10^{14.5}$
(open squares and dashed lines respectively).  Lower mass curves have
been off-set vertically by 1 dex for clarity.  The number of
contributing haloes for the lower mass range is $\approx$ 41 $\times 10^3$ and 12
$\times 10^3$ (for increasing redshifts), for the higher mass range 
19$\times 10^3$ and 0.5$\times 10^3$}
\end{figure*}

We have applied \pin\ to the initial conditions of a simulation by
Governato et al (1999). This large volume dissipationless simulation
uses $360^3$ dark matter particles and was evolved using the PKDGRAV
Tree code (comoving box size 500 $h^{-1}$ Mpc, matter density
$\Omega_m=1$, Hubble constant $H_0=50$ km s$^{-1}$ Mpc$^{-1}$, standard
CDM spectrum with $\sigma_8=1$).  Haloes have been identified at
several output times using a standard FOF algorithm with linking-length
$b=0.2$.  \pin\ is fast: resampling the initial conditions onto a
256$^3$ grid, the first stage of computing orbit-crossing requires
$\sim$ 6 hours of CPU time, the second step of identifying the haloes
takes just a few minutes. (Timings refer to a Pentium~III 450MHz
personal computer. Memory requirement in this case amounts to $\sim
512$Mbytes of RAM.)  These timings should be
contrasted with the several hundreds of hours on 256 nodes of a T3E
Cray supercomputer required to perform the original simulation.
Moreover, \pin\ immediately outputs the merger tree of each halo, which
should be contrasted to the complicated and expensive postprocessing
necessary to extract merger trees from a simulation.

One way to understand the large speed-up between an $N$-body simulation
and \pin\ is that most of the CPU time used in the $N$-body simulation
is spent integrating the orbits for particles {\em already inside a
halo}. These particles undergo large accelerations as they orbit inside
the halo, and hence may require thousands of time-steps in order for
their orbits to be integrated accurately. \pin\, on the other hand,
completely ignores particles once they are inside a halo and so can use
far fewer steps per particle to perform the whole simulation, since it
only needs to compute the particle's orbit {\em before} it enters any
high density region. Obviously, all information on the internal
structure of the halo is lost in the process, but it is well known that
several millions of particles are required to get the internal
structure correct. (See the controversy about the slope of halo
profiles as determined using high-resolution collisionless simulations,
e.g. Ghingha et al. 2000).
In the following we
demonstrate that \pin\ is indeed able to
predict the merging and clustering properties of haloes very
accurately.

\pin\ reproduces the mass function $M n(M)$ (number of objects per unit
volume and unit $\ln M$) to better than 10 per cent at all redshifts
(Figure~\ref{fig:mfunc}), in the mass range in which haloes have at
least $\sim$30 particles and Poisson error bars are small.  
To make this more evident, we plot in the lower panel of
Figure~\ref{fig:mfunc} the residuals with respect to the $z=0$ FOF
mass function.
This level of accuracy improves over the fit proposed by Sheth and
Tormen (2000). (The PS mass function, which over (under) predicts the
number of low (high) mass objects, is shown for comparison as well.)

The good agreement between halo masses is not just statistical in
nature. We have plotted in figure~\ref{fig:dplot} the masses of the
haloes that a particle is assigned to, for both \pin\ and FOF haloes,
for a random subset of particles drawn from the initial conditions.
The correlation between \pin\ and FOF masses is extremely tight, and is
dramatically better than PS (compare with Figure~2 in Sheth, Mo \&
Tormen 2001), and also improves over peak-patch (Bond \& Myers 1996b,
their figure~11).

Figure~\ref{fig:dplot} contains outliers which correspond to particles
that are assigned to a different halo (or are not assigned to a halo at
all) by \pin\ than by the simulation (or vice versa). We have
investigated in detail the typical overlap in the initial conditions
between simulated haloes and those found by \pin. Since \pin\ refers to
the same initial conditions as the simulation, we can determine the
fraction of Lagrangian volume $V_{\rm P}$ of a given halo identified by
\pin\ that overlaps the Lagrangian volume $V_{\rm FOF}$ of a FOF
halo. In general, for any FOF halo, the volume $V_{\rm FOF}$ may
overlap with the Lagrangian volumes of several \pin\ haloes (and vice
versa). For example, if two \pin\ haloes fail to merge, whereas the
corresponding FOF haloes do merge, then the volume $V_{\rm FOF}$ may be
broken-up into two \pin\ volumes $V_{\rm P}$.  We choose to pair-up two
haloes between the two catalogs, if their Lagrangian volumes overlap to
better than 30 per cent. Paired-up haloes are in addition called \lq
cleanly assigned\rq\, if the intersection of $V_{\rm FOF}$ with $V_{\rm
P}$ is larger than for any other FOF halo, and vice versa. Paired-up
haloes that are not cleanly assigned are called \lq split\rq. Denoting
the fraction of haloes that are cleanly assigned and the fraction that
are split by $f_{\rm cl}$ and $f_{\rm split}$ respectively, then the
fraction of haloes that are not paired-up is obviously $1-f_{\rm
cl}-f_{\rm split}$.

In figure~\ref{fig:obj} we show $f_{\rm cl}$ and $f_{\rm split}$ as a
function of halo mass, for several redshifts.  For sufficiently
massive haloes $M\ge 10^{14} M_\odot$ (corresponding to 40 particles),
$f_{\rm cl} \ge 0.8$, showing that most FOF haloes can be unambigously
associated with a corresponding \pin\ halo, while the fraction of FOF
haloes split in two or more \pin\ haloes is small. The fraction
$1-f_{\rm cl}-f_{\rm split}$ of FOF haloes that have no corresponding
\pin\ halo is very small as well, ranging from $\la 1$ per cent for
the most massive haloes, to $\sim 15$ per cent for small haloes with
$\sim$40 particles. The latter limit is close to the minimum number
needed to correctly numerically simulate the formation of a halo,
given an initial density field. For cleanly assigned haloes, the
bottom panel in the figure shows the fractional overlap $f_{\rm ov}$
of the respective Lagrangian volumes. A typical values for well
resolved haloes is $f_{\rm ov}\sim 0.7$, indicating that the mass
errors are usually smaller than 30 per cent. This is made more clear
in figure ~\ref{fig:dplot2}, which compares the FOF with the \pin\
masses for cleanly assigned halo pairs.  The correlation is very
tight. The level of agreement between \pin\ and simulations is only
weakly dependent on redshift.

Since \pin\ haloes are in detail very similar to their corresponding
FOF haloes, their merging history and clustering properties can be
expected to be very similar as well. The conditional mass function
$n(M,z|M_0,z_0)$ (the number density of objects of mass $M$ at redshift
$z$ that are merged in haloes of mass $M_0$ at the later redshift
$z_0$) is shown in Figure~\ref{fig:prog}. The PS prediction, computed
following Bond et al. (1991; see also Bower 1991; Lacey \& Cole 1993)
is also shown.  Also in this case the agreement between \pin\ and the
simulation is very good, making an improvement with respect to PS and
demonstrating that \pin\ haloes undergo a very similar merging history
as do FOF haloes.

Finally, we compare in Figure~\ref{fig:corr} the two-point correlation
function $\xi(r)$ of haloes as a function of mass and redshift. The
agreement with the simulations is again very good.  In particular, the
high clustering amplitude of massive haloes at early times is well
reproduced, and the correlation length $r_0$ is recovered to within 10
per cent or better, thus improving the PS-like estimate of Sheth et
al. (2001) and allowing to discriminate easily between different
cosmological models (Colberg et al. 2001).
The quality of this agreement suggests that halo positions are well
estimated by \pin; we find that the 1D rms error on the final positions
is $\sim$ 0.8 $h^{-1}$ Mpc (smaller than the grid spacing), while
velocities are recovered with a 1D rms of $\sim$150 km/s.

\section{Conclusions}
We have demonstrated that \pin\ is able to accurately describe the
evolution of clustering of haloes as a function of mass. Therefore,
when combined with semi-analytical models for galaxy formation (White
\& Frenk 1991, Kauffmann, White \& Guiderdoni 1993, Cole et al 1994,
Somerville \& Primack 1999), \pin\ can be used to reliably generate
mock galaxy catalogs, with the correct evolution of galaxy clustering
build-in, while requiring orders of magnitude less computer time than
numerical simulations. Easy and accurate production of large halo
catalogues is invaluable for interpreting data and estimating errors
from galaxy or galaxy cluster surveys, for example when studying
galaxy bias (Diaferio et al. 1999; Benson et al. 2000), estimating
power-spectra (e.g. Efstathiou \& Moody 2001), determining shear from
weak lensing measurements (van Waerbeke et al. 2000, Wittman et
al. 2000, Bacon, Refregier \& Ellis 2000, Kaiser, Wilson \& Luppino
2000) or studying intrinsic galaxy alignments (Crittenden et al. 2001,
Brown et al. 2000).

A more detailed and technical account account of the code, suitable
for those who wish to use it, will be given in a forthcoming paper
(Monaco et al. 2001), while the ability of predicting halo merger
histories beyond the progenitor mass function will be presented by
Taffoni et al. (2001).  A public version of \pin\ is available at the
site http://www.daut.univ.trieste.it/pinocchio.

\section*{Acknowledgements}
We thank Jasjeet Bagla, Stefano Borgani, Anatoly Klipin, Barbara
Lanzoni, Sergei Shandarin and Ravi Sheth for many discussions.
$N$-body simulations were run at the ARSC and Pittsburg supercomputing
centers. PM acknowledges support from MURST.  TT acknowledges support
from the \lq Formation and Evolution of Galaxies\rq\ network set up by
the European Commission under contract ERB FMRX-CT96086 of its TMR
programme, and from PPARC for the award of of post-doctoral
fellowship.  Research conducted in cooperation with Silicon
Graphics/Cray Research utilising the Origin 2000 super computer at
DAMTP, Cambridge.

{}
\end{document}